\documentclass[11pt]{article}
\pdfoutput=1

\usepackage{euscript}
\usepackage{amssymb}
\usepackage{amsfonts}
\usepackage{amsbsy}
\usepackage{epsfig}
\usepackage{amsthm}
\usepackage{amscd}
\usepackage{amstext}
\usepackage{verbatim}
\usepackage{amsmath}
\usepackage{cancel}
\usepackage{capt-of}
\usepackage{empheq}
\usepackage{subfigure}

\usepackage[pdftex]{hyperref}

\def\ben{\begin{equation}}
\def\een{\end{equation}}
\def\half{{\textstyle{1\over2}}}

\def\be{\begin{equation}}
\def\ee{\end{equation}}
\def\beq{\begin{equation}}
\def\eeq{\end{equation}}
\def\ba{\begin{array}}
\def\ea{\end{array}}

\def\dalemb#1#2{{\vbox{\hrule height .#2pt
       \hbox{\vrule width.#2pt height#1pt \kern#1pt
               \vrule width.#2pt}
       \hrule height.#2pt}}}

\newcommand{\bea}{\begin{eqnarray}}
\newcommand{\eea}{\end{eqnarray}}

\def\ocal{{\mathcal{O}}}

\begin{document}

\begin{center}

{ \Large {\bf
Disordered horizons: \\
Holography of randomly disordered fixed points
}}

\vspace{1cm}

Sean A. Hartnoll$^{1}$ and Jorge E. Santos$^{1,2}$

\vspace{1cm}

{\small
$^{1}${\it Department of Physics, Stanford University, \\
Stanford, CA 94305-4060, USA }}

\vspace{0.5cm}

{\small
$^{2}${\it Department of Applied Mathematics and Theoretical Physics, \\
University of Cambridge, Wilberforce Road, \\
Cambridge CB3 0WA, UK}}

\vspace{1.6cm}

\end{center}

\begin{abstract}

We deform conformal field theories with classical gravity duals by marginally relevant random disorder. We show that the disorder generates a flow to IR fixed points with a finite amount of disorder. The randomly disordered fixed points are characterized by a dynamical critical exponent $z>1$ that we obtain both analytically (via resummed perturbation theory) and numerically (via a full simulation of the disorder). The IR dynamical critical exponent increases with the magnitude of disorder, probably tending to $z \to \infty$ in the limit of infinite disorder.

\end{abstract}

\pagebreak
\setcounter{page}{1}

\section{Introduction}

The effects of random disorder in quantum critical systems remains a challenging problem, especially in spatial dimension greater than one. Quenched random disorder is introduced into a critical theory as a coupling in the theory that is spatially inhomogeneous, with the spatial variation drawn from a random distribution. As with the usual homogeneous couplings at a critical point, the most important fact to determine is whether the random coupling is relevant or irrelevant. A simple generalization of the standard power-counting arguments can be applied to random couplings -- as we review below -- and the result is called the Harris criterion \cite{harris}. If the disorder satisfies the Harris criterion then it is irrelevant and can be treated perturbatively at low energy scales. If the disorder is relevant, however, then the low energy physics has the potential to be very complicated. Solvable models have indicated two types of behavior in this case: either the system flows to a new fixed point in which the disorder takes a finite value, see e.g. \cite{Weinrib:1983zz} for a statistical mechanics example, or the disorder can grow without bound leading to `infinite randomness' fixed points. See e.g. \cite{sachdev, mot, vojta} for discussions of this case and further references.

Holography is a powerful framework for describing the physics of strongly interacting quantum field theories in general, and quantum critical points in particular. It is a strongly-interacting large $N$ limit (in contrast to the typically weakly interacting fixed points that can be accessed via a vector large $N$ limit) that captures quantum field theoretic physics in classical solutions to general relativity with a negative cosmological constant. The description of disorder has remained challenging, however, because general relativity is a highly nonlinear classical theory and finding solutions in the presence of a complicated disordered source is daunting. Nonetheless, holographic studies have successfully computed a key effect due to disorder: disorder relaxes momentum and hence leads to finite electrical conductivities in systems with a net charge \cite{Hartnoll:2007ih, Hartnoll:2008hs, Hartnoll:2012rj, Anantua:2012nj,Lucas:2014zea}. This has been achieved for cases in which disorder can be treated perturbatively: at low temperatures for irrelevant disorder or at high temperatures for relevant disorder. Disorder remains important in these calculations because, while irrelevant or otherwise small, it provides the leading contribution to momentum relaxation.

In this work we will present the first holographic solutions with inherently disordered zero temperature near horizon geometries. These will describe the disordered IR fixed points of (marginally) relevant disorder about a UV critical theory. Our approach to finding the solutions is similar in spirit to that in \cite{Adams:2011rj, Adams:2012yi}. Namely, we perturbatively implement a disorder averaging procedure on solutions to the Einstein equations. The difference will be that we are able to resum the perturbation theory, in a certain limit, to obtain a small but finite effect of the disorder in the far IR geometry where the original perturbation theory would break down due to the growth of the disordered coupling. Thus we obtain a genuinely disordered horizon.

Our main result is that for marginally relevant random deformations of the UV conformal field theory there exists an IR fixed point with dynamical critical exponent $\overline z$ greater than one.\footnote{We put a bar over the dynamical critical exponent $\overline z$ in order to distinguish it from the holographic coordinate $z$ that we introduce shortly.} The UV fixed point in contrast has $\overline z =1$. In particular, the averaged near horizon geometry is a Lifshitz horizon \cite{Kachru:2008yh}. The dynamical exponent increases with the strength of disorder. Perturbatively in the strength $\bar V$ of the disorder we obtain analytically
\be\label{eq:zgen}
\overline z = 1+\frac{\pi^{\frac{D-3}{2}}}{2}\Gamma\left(\frac{D-1}{2}\right)\,\bar{V}^2+\mathcal{O}(\bar{V}^4)\,.
\ee
Here $D$ is the number of bulk spacetime dimensions, $\Gamma$ is the gamma function and the bulk theory is AdS$_D$ gravity coupled to a massive scalar field. The scalar field is dual to an operator with a random disordered coupling and that saturates the Harris criterion. For the case of three bulk dimensions, we are able to go to higher order in perturbation theory and obtain
\begin{equation}\label{eq:zthree}
\overline z = 1+\half \bar{V}^2 + \half (\ln 2)\,\bar{V}^4+\mathcal{O}(\bar{V}^6)\,.
\end{equation}
We corroborate both of these expressions numerically and furthermore are able to numerically establish the presence of an IR Lifshitz horizon in the three dimensional case up to $\bar V \approx 2$, giving $\overline z \approx 12$. In obtaining these results it is important to solve the bulk gravitational equations of motion exactly, in the presence of a random boundary source, and only afterwards perform the disorder average.

The results in the previous paragraph establish the existence of a holographic IR fixed point with finite randomness for the case of a marginally relevant UV deformation. For an irrelevant deformation, satisfying the Harris criterion, the geometry will clearly flow back to $\overline z=1$ in the IR. For a fully relevant UV deformation, that violates the Harris criterion, a strong effect on the IR is expected \cite{us}. However, a simple possibility for the IR geometry is that it may be described by an extremal horizon with $\overline z = \infty$! This would give an attractive picture in which $\overline z$ is tuned from $1$ to $\infty$ by marginally relevant disorder separating the irrelevant ($\overline z=1$) and relevant ($\overline z = \infty$) cases. Remarkably, solvable lattice models with infinite disorder fixed points show an emergent $\overline z = \infty$ \cite{danfisher}.

Stable disordered quantum critical systems with finite ${\overline z}$, of the type we find, have proved elusive in conventional Wilson-Fisher theory. There, disordered fixed points are unstable against growth of disorder \cite{ykwen}, unless a double $\varepsilon$-expansion is used in which the number of time dimensions is close to zero \cite{cardy}. This expansion connects with a stable statistical mechanics fixed point \cite{Weinrib:1983zz}, but is of unclear physical significance. See \cite{sachdev} for a discussion.

\section{\label{sec:setup}Setup}

To obtain the desired physics in the simplest setting possible we will work in 2+1 bulk dimensions for most of this paper. We will also present results for a general $D$ dimensional bulk at the end. The minimal ingredients of our holographic model are the usual Einstein-Hilbert term and a neutral scalar field, which we will use to source disorder. Our starting point is the following three dimensional bulk action
\begin{equation}
S = \frac{1}{16\pi G_N}\int \mathrm{d}^3 x\,\sqrt{-g}\left[R+\frac{2}{L^2}-2\nabla_a \Phi \nabla^a \Phi-4 V(\Phi)\right]\,,
\label{eq:action}
\end{equation}
where $L$ is the AdS length and $G_N$ is Newton's constant. We will take a free scalar field with a negative mass squared,
so that
\begin{equation}
V(\Phi) = -\frac{\mu\,\Phi^2}{2 L^2}\,.
\label{eq:potential}
\end{equation}
The equations of motion derived from (\ref{eq:action}) take the following form
\begin{subequations}
\begin{align}
&R_{ab} = 2\left[\nabla_a \Phi \nabla_b \Phi + 2 V(\Phi)g_{ab}\right]-\frac{2}{L^2}g_{ab} \,,
\\
&\Box \Phi -V^\prime(\Phi)=0\,.
\end{align}
\end{subequations}

We are interested in solutions to these equations that asymptote to the Poincar\'e patch of $AdS_3$. That is
\be\label{eq:boundarymetric}
ds^2 = \frac{L^2}{z^2} \left( \mathrm{d}z^2 -\mathrm{d}t^2+\mathrm{d}x^2  + \cdots \right) \,,
\ee
as $z \to 0$. We take the scalar field to have $\mu = 3/4$; we will see shortly that this corresponds to an operator that saturates the Harris criterion in the dual field theory. In this case $\Phi$ takes the following form near the conformal boundary as $z \to 0$:
\begin{equation}
\Phi = z^{1/2}\,\Phi_1(x)+z^{3/2}\,\Phi_2(x)+\cdots\,. \label{eq:phiasymptotic}
\end{equation}
As usual in the AdS/CFT correspondence \cite{Maldacena:1997re,Witten:1998qj,Gubser:1998bc}, $\Phi_1$ is identified as the source to the operator $\mathcal{O}$ dual to $\Phi$, whereas $\Phi_2$ is its expectation value $\langle \ocal \rangle$.

Throughout this paper we will realize disorder using a spectral representation, following \cite{Shinozuka1991}. We briefly review the essentials of this method. The disordered source is written as
\begin{equation}
\Phi_1(x) = \bar V \sum_{n=1}^{N-1} A_n\,\cos(k_n\,x+\gamma_n)\, .
\label{eq:randompot}
\end{equation}
Here the frequencies are evenly spaced: $k_n = n \Delta k$, with $\Delta k = k_0/N$. The amplitudes are:
\begin{equation}
A_n =  2\,\sqrt{S(k_n)\Delta k}\,. \label{eq:amplitude}
\end{equation}
The $\gamma_n$ are random phases uniformly distributed in $(0,2\pi)$. The shortest periodicity of the potential is (in the large $N$ limit) $x \sim x + 2\pi/k_0$ and hence $1/k_0$ defines a short distance cutoff to the disorder. This cutoff is kept fixed in the limit $N\to \infty$, so that the spacing between frequencies $\Delta k \to 0$ in this limit. The largest periodicity is $x \sim x + 2\pi N /k_0$, and so $N/k_0$ is an infrared cutoff on the disorder. Finally, the function $S(k)$ appearing in the amplitude (\ref{eq:amplitude}) controls the correlations in the noise and will be chosen shortly.

We denote the random average of any quantity $f$ by $\langle f\rangle_R$, which is defined by the limit
\begin{equation}
\langle f \rangle_R = \lim_{N\to+\infty} \int \prod_{i=1}^{N-1}  \frac{\mathrm{d}\gamma_i}{(2\pi)} f\,.
\end{equation}
In this limit, the potential (\ref{eq:randompot}) describes Gaussian noise \cite{Shinozuka1991}. If we further make the choice $S(k) = 1$, then it describes local Gaussian noise, so that
\begin{equation}
\langle \Phi_1(x) \rangle_R = 0 \,, \qquad
\langle \Phi_1(x)\Phi_1(y)\rangle_R = \bar{V}^2 \delta(x-y)\,,
\label{eq:gaussiannoise}
\end{equation}
with all higher correlators determined by Wick contractions. We will stick with the choice $S(k) = 1$
throughout this paper, even though our calculations can be easily adapted to other functions that result in disorder with longer range correlation.\footnote{If the short distance cutoff $1/k_0$ is retained on the disorder distribution (\ref{eq:randompot}) as $N \to \infty$, as it will be throughout this paper, then very high energy probes of the disorder will see oscillations at the large frequency $k_0$ superimposed onto the delta function distribution in (\ref{eq:gaussiannoise}).}

The spectral approach to disorder has recently been used in a holographic setting \cite{Arean:2013mta}.

The Harris criterion is easily derived as follows. The coupling of the dual field theory to the disordered source is
\be
\int \mathrm{d}t\mathrm{d}x \, \Phi_1(x) \ocal(t,x) \,.
\ee
Let the mass dimension of the operator be $[\ocal] = \Delta$. Then clearly $[\Phi_1] = 2 - \Delta$. The strength of the disorder is however controlled by $\bar V$ in (\ref{eq:gaussiannoise}). From (\ref{eq:gaussiannoise}) we have $[\bar V] = [\Phi_1] - 1/2 = 3/2 - \Delta$. It follows that disorder is relevant if $\Delta < 3/2$, irrelevant for $\Delta > 3/2$ and marginal for $\Delta = 3/2$. We now see that the choice of mass in the bulk leading to (\ref{eq:phiasymptotic}) indeed corresponds to marginal disorder. This scaling argument is immediately generalized to $D$ bulk spacetime dimensions to give that the disorder is irrelevant for $\Delta > D/2$.

\section{\label{sec:analytic}Perturbation theory and resummation thereof}

This section develops a bulk perturbation theory in the strength of disorder. The following section will present numerical simulations that, among other things, will confirm our resummed perturbative results. The procedure employed is very similar to the one used in \cite{Dias:2011ss,Dias:2012tq}. We start by writing the most general line element compatible with the symmetries of our problem. Since we are interested in zero temperature horizons, there will be a static Killing vector field $\partial_t$. Furthermore, in this section, we adopt the Fefferman-Graham gauge, which brings our line element and neutral scalar field to the following simple forms
\begin{equation}
\mathrm{d}s^2 = \frac{L^2}{z^2}\left[-A(x,z)\mathrm{d}t^2+B(x,z)\mathrm{d}x^2+\mathrm{d}z^2\right],\quad\text{and}\quad\Phi = \widehat{\Phi}(x,z)\,.
\label{eq:perturbativeansatz}
\end{equation}
Our objective is then to solve for the three functions of two variables $A$, $B$ and $\widehat{\Phi}$.

Since the stress energy tensor is even in the scalar field $\Phi$, the following expansion for $\bar{V}\ll1$ holds:
\begin{equation}
A =1+\sum_{i=1}^{\infty}\bar{V}^{2i} A^{(2i)}(x,z)\,,\quad B =1+\sum_{i=1}^{\infty}\bar{V}^{2i} B^{(2i)}(x,z) \,, \quad\widehat{\Phi} = \sum_{i=0}^{\infty} \bar V^{2i+1}\widehat{\Phi}^{(2i+1)}(x,z)\,.
\label{eq:expansion}
\end{equation}
If $\bar{V}=0$, this is an exact solution of the Einstein scalar equations, namely AdS$_3$ written in Poincar\'e coordinates. We then proceed our calculation in a power series in $\bar{V}$. At first order we find that the required solution, whose source is given in Eq.~(\ref{eq:randompot}), is simply
\begin{equation}
\widehat{\Phi}^{(1)}(x,z)=\sqrt{z} \sum_{j=1}^{N-1}A_j\,\cos(k_j\,x+\gamma_j)\,e^{-k_j\,z}\,.
\label{eq:linearesult}
\end{equation}
Even though this is a linear result, it already has some interesting consequences. The expectation value of the operator dual to $\Phi$ is seen to be
\begin{equation}
\langle \mathcal{O}^{(1)}(x)\rangle =- \bar V \sum_{j=1}^{N-1}A_j\,k_j\,\cos(k_j\,x+\gamma_j)\,.
\end{equation}
This quantity has a different statistics from the source. With our choice of $A_j = \text{const.}$, then $\langle \langle \mathcal{O}^{(1)}_{\Phi}(x)\rangle \langle \mathcal{O}^{(1)}_{\Phi}(y)\rangle\rangle_{R} \propto \delta^{\prime\prime}(x-y)$. This is compatible with
the numerical observation made in \cite{Arean:2013mta} that several responses were more irregular than the sources.

The important step is to feed the linear result (\ref{eq:linearesult}) back into the Einstein equations to find the disordered metric at second order in perturbation theory. We do this in appendix \ref{sec:second}. The computation can be done analytically. While the metric expressions at second order are complicated, after disorder averaging, the solution simplifies dramatically to the following:
\begin{align}
&\langle A^{(2)}(x,z) \rangle_{R} =e^{-2 k_0 z} \left\{(1+2 k_0 z) \left[\ln (2 k_0 z)+\gamma \right]-2 k_0 z\right\}-\ln (2 k_0 z)\,,
\label{eq:limitcentral}
\\
&\langle B^{(2)}(x,z)\rangle_{R} =e^{-2 k_0 z}+\gamma -1\,,\nonumber
\end{align}
where $\gamma$ is Euler's constant. Deep in the IR, where $z\to\infty$, $\langle A^{(2)}(x,z) \rangle_{R}$ diverges logarithmically, signalling that the disorder is in fact marginally relevant. This is very similar to the marginally relevant holographic disorder found in \cite{Adams:2012yi}. The logarithmic divergence clearly indicates a breakdown of perturbation theory.

In order to restore the validity of perturbation theory, we need to resum the logarithmic divergence. This is easily achieved by a Poincar\'e-Lindstedt resummation technique. In our context, this is tantamount to changing the line element (\ref{eq:perturbativeansatz}) to
\begin{equation}
\mathrm{d}s^2 = \frac{L^2}{z^2}\left[-\frac{A(x,z)\,\mathrm{d}t^2}{F(z)^{\beta(\bar{V})}}+B(x,z)\mathrm{d}x^2+\mathrm{d}z^2\right]\,,
\end{equation}
where $F(z)$ is a function of our choice that must be $1$ at $z=0$ and diverge at large $z$. Furthermore, $\beta(\bar{V})$ is a function of $\bar{V}$ only, which we expand as:
\begin{equation}
\beta(\bar{V}) = \sum_{i=1}^{+\infty}\bar{V}^{2i}\beta_{2i}\,.
\end{equation}
Different choices of $F$ will correspond to different resummations, all of which lead to the same IR geometry. A convenient choice that eases our calculations at higher order in $\bar{V}$ is:
\begin{equation}
F(z) =1+k_0^2\,z^2\,\quad\text{and}\quad \beta_2 = \frac{1}{2}\,.
\end{equation}

One can redo our perturbation scheme outlined above \emph{mutatis mutandis}, to find that
\begin{equation}
\langle A^{(2)}(x,z) \rangle_{R} = e^{-2 k_0 z} \left\{(1+2 k_0 z) \left[\ln (2 k_0 z)+\gamma \right]-2 k_0 z\right\}-\ln \left(\frac{2 k_0 z}{\sqrt{1+k_0^2\,z^2}}\right)\,,
\end{equation}
with $\langle B^{(2)}(x,z) \rangle_{R}$ again given by the second equation in (\ref{eq:limitcentral}). Note that $|\langle A^{(2)}(x,z) \rangle_{R}|$ is a bounded function, never larger than the background, over the entire range of $z$. This process has, however, changed the behavior of $g_{tt}$ in the deep IR. As $z \to \infty$ the spacetime now takes the Lifshitz \cite{Kachru:2008yh} form. That is, rescaling $t$ and $x$ to absorb constants terms,
\be
\mathrm{d}s_\text{IR}^2 = \frac{L^2}{z^2}\left(-\frac{\mathrm{d}t^2}{z^{2 \beta(\bar V)}}+ \mathrm{d}x^2+\mathrm{d}z^2\right) \,.
\ee
The dynamical critical exponent is $\overline z = 1 + \beta(\bar V)$.

We have carried out this analysis to fourth order in perturbation theory, and the procedure outlined above still carries through. In particular we see no breakdown of perturbation theory after the resummation. The resulting metric and scalar field are too cumbersome to be presented here, so we just quote the final result. To fourth order we find that the IR geometry is accurately described by a Lifshitz metric, with exponent
\begin{equation}
\overline z = 1+ \half \bar{V}^2+ \half (\ln 2)\,\bar{V}^4+\mathcal{O}(\bar{V}^6)\,.
\label{eq:central}
\end{equation}
This is the result given in the introduction.

It is straightforward to generalize the second order calculation to an arbitrary number of bulk spacetime dimensions $D$. We take the same action (\ref{eq:action}) for gravity plus a massive scalar, but in $D$ dimensions. The Harris criterion is saturated by $\Delta = D/2$, using standard boundary conditions for the scalar. The dynamical critical exponent is found to be
\begin{equation}\label{eq:genD}
\overline z = 1+\frac{\pi^{\frac{D-3}{2}}}{2}\Gamma\left(\frac{D-1}{2}\right)\,\bar{V}^2+\mathcal{O}(\bar{V}^4)\,,
\end{equation}
where $\Gamma$ is the Gamma function. Note that if we set $D=3$, we readily reproduce Eq.~(\ref{eq:central}) to second order in $\bar{V}$. This result was also given in the introduction. We outline the calculation leading to (\ref{eq:genD}) in appendix \ref{sec:generalD}.

\section{\label{sec:num}Numerics}

This section considers the effects of disorder in regimes where $\bar{V}$ is order unity. At present, this problem does not seem amenable to analytic calculations, so we proceed with a direct numerical integration of the Einstein equations.

We are interested in solutions with a zero temperature horizon. This not only means that there is one time-like Killing vector field $\partial_t$, but also that the norm of $\partial_t$ must vanish at least quadratically at the extremal horizon. Also, close to the conformal boundary, we demand that our line element approach AdS in Poincar\'e coordinates, Eq.~(\ref{eq:boundarymetric}). We again start with the case of $D=3$, and so the most general line element compatible with such requirements reads
\begin{equation}
\mathrm{d}s^2 = \frac{L^2}{r^2}\left\{(1-r)^2\left[-A\, \mathrm{d}t^2+S\,(\mathrm{d}x+F\,\mathrm{d}r)^2\right]+\frac{B\,\mathrm{d}r^2}{(1-r)^2}\right\}\quad\text{and}\quad \Phi = L\,\sqrt{\frac{r}{1-r}}\,P \,,
\label{eq:lineelementgeneral}
\end{equation}
where $A$, $B$, $F$, $S$ and $P$ are five functions of two variables, $x$ and $r$, to be determined by our numerical scheme. Here henceforth, we set the AdS lengthscale $L = 1$. Furthermore, $r=0$ denotes the conformal boundary and $r=1$ the Poincar\'e horizon. Our metric \emph{ansatz} (\ref{eq:lineelementgeneral}) is such that when $A=B=S=1$ and $P=F=0$, we recover the usual metric of AdS$_3$ written in Poincar\'e coordinates with $z = r/(1-r)$ being the usual Fefferman-Graham holographic direction. We have also stripped off a power $r^{1/2}$ from $\Phi$, so that, at $r=0$, $P$ is given by the source (\ref{eq:randompot}).

The numerical scheme we use in this manuscript is similar to the one used in \cite{Horowitz:2012ky}, which was first introduced in \cite{Headrick:2009pv} and studied in great detail in \cite{Figueras:2011va}. We shall only briefly detail the main differences and difficulties. First, we are interested in zero temperature horizons. This means that, at the Poincar\'e horizon, there will be some non-analytic behavior that is not well adapted to spectral methods\footnote{To be precise, if we Fourier decompose the scalar field at infinity as $\Phi(x,z) = \sum_n a_n(z)\,\cos(k_n x+\gamma_n)$, the metric functions approach a constant value at the Poincar\'e horizon as $\exp[-2 k_n/(1-r)]$ and powers thereof, which corresponds to an essential singularity of type $I$.}. Furthermore, in $D=3$, there is a logarithmic behavior in $P$ close to $r=0$, which is also a conundrum for spectral methods. In order to bypass this, we patch two finite difference grids on to the spectral grid (one close to the boundary, the other close to the Poincar\'e horizon). If these patching regions are not too large, the numerical method still exhibits exponential convergence. Furthermore, in order to help convergence, we monitor the gradients close to the boundary and horizon, and whenever needed, we double the number of finite difference patches in order to ensure the desired accuracy, \emph{i.e.} we use adaptive mesh refinement. Finally, in the region in $r$ where spectral collocation methods are used, we take a Chebyshev spectral collocation basis, and in $x$ we use the Fourier nodes.

A couple of comments are in order regarding the simulation of the random potential (\ref{eq:randompot}). Even for small $\bar{V}$, the pointwise value of $P(0,x)$ increases with $\sqrt{N}$. This means that for all simulations we have to use a very large number of grid points, and that the larger $N$, the more complicated it is to extract the relevant physical quantities accurately. Of course, the problem gets even more severe when $\bar{V}$ increases. For this reason, the error bars in our computations increase with $\bar{V}$. Also, for higher-dimensional calculations, say in $D=4$, this problem is far more serious, because the pointwise value of $P(0,x,y)$ now increases with $N$.

Finally, we describe how to extract the dynamical scaling exponent ${\overline z}$ in a gauge invariant way. Any pure $D-$dimensional Lifshitz geometry can be written as
\begin{equation}
\mathrm{d}s_\text{Li}^2 = \frac{1}{z^2}\left[-\frac{\mathrm{d}t^2}{z^{2 ({\overline z}-1)}} + \sum_{i=1}^{D-2}\mathrm{d}x_i^2+\mathrm{d}z^2\right]\,,
\end{equation}
where lower latin letters run over the boundary spatial directions. In these coordinates, the norm of the Killing field $\partial_t$ is $\ell \equiv \sqrt{-\partial_t\cdot\partial_t}= z^{-{\overline z}}$. Furthermore, slices of constant $t$ and constant $\ell$ define a $(D-2)-$dimensional torus, whose volume scales as
\begin{equation}
\mathcal{V}_{\mathbb{T}^{D-2}} = \ell^{\frac{D-2}{{\overline z}}} \,.
\end{equation}
Both left and right hand sides of this equation are coordinate invariant. From this scaling we can extract the numerical value of ${\overline z}$, by fitting the data to the relevant range in $\ell$.

Another way to determine ${\overline z}$ in Fefferman-Graham coordinates is via a logarithmic derivative. Define:
\begin{equation}\label{eq:logderiv}
{\overline z}(z) \equiv - \frac{1}{2} \,\frac{z}{\langle g_{tt} \rangle_R} \frac{d \langle g_{tt} \rangle_R}{dz}\,.
\end{equation}
In a Lifshitz scaling regime this quantity will be constant.
We have explicitly checked that both diagnostics given above agree with each other, the second one being much easier to evaluate. For this reason, we use the second diagnostic in what follows.

\begin{figure}[h]
\centering
\subfigure{\label{fig:0a} (a) \includegraphics[height = 0.26\textheight]{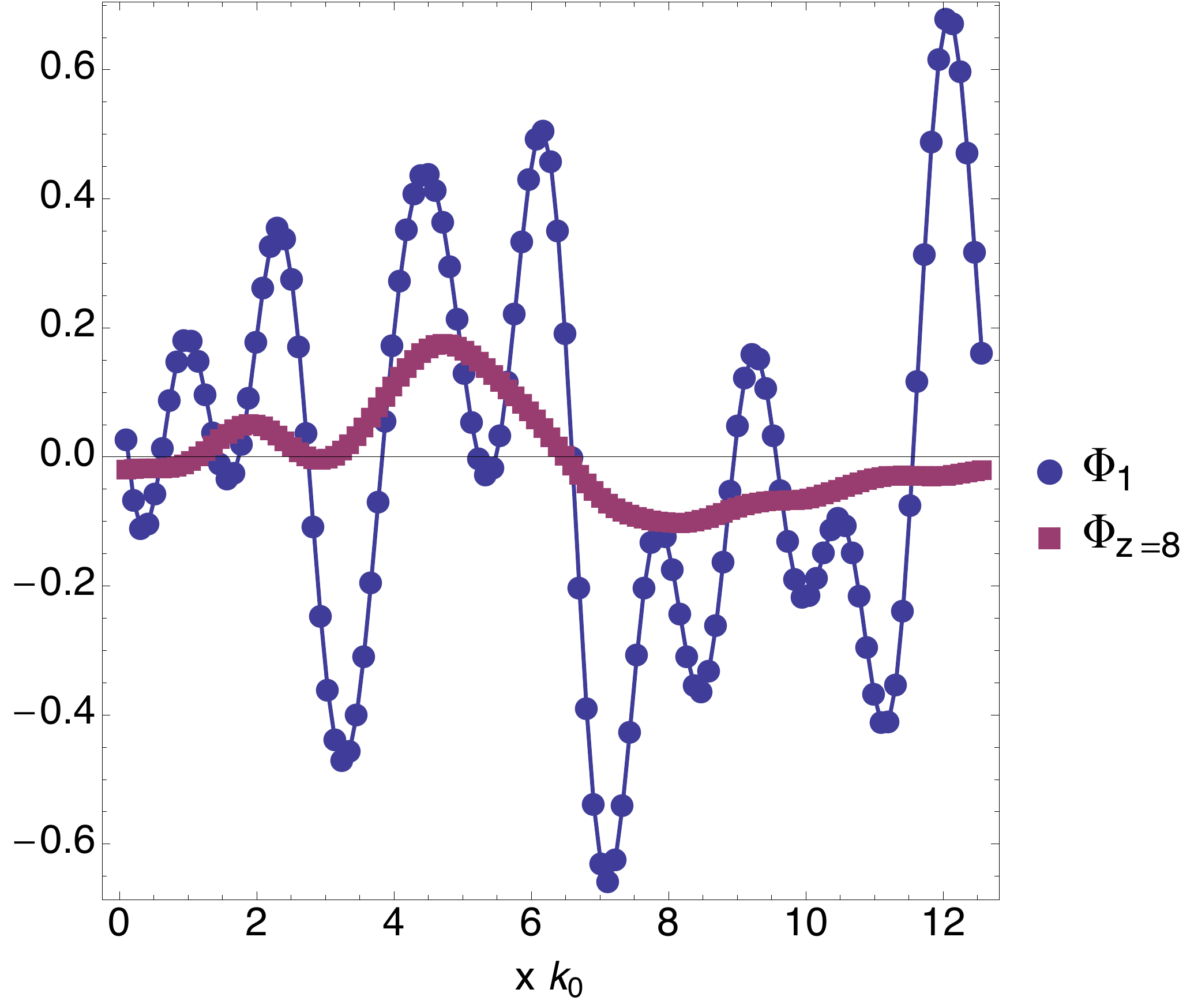}}
\subfigure{\label{fig:0b} (b) \includegraphics[height = 0.26\textheight]{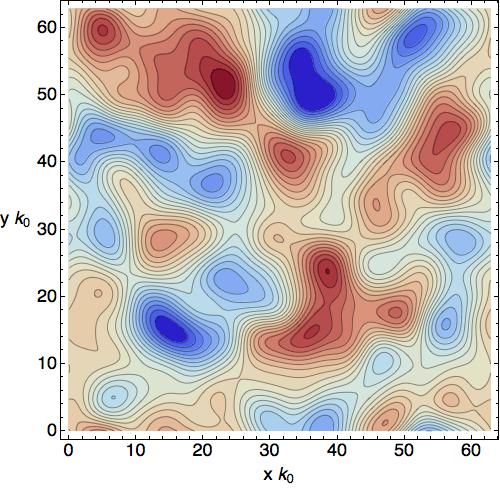}}
\caption{\label{figs:0} {\bf Disordered horizons:} Plot (a) shows the scalar source $\Phi_1$ as a function of $x\,k_0$ at the boundary and the scalar field $\Phi$ at $z=8$, which is well within the IR Lifshitz scaling regime. Plot (b) is a density plot of $\Phi$, now in a $D=4$ spacetime, for $\ell = 1$, as a function of boundary directions $x\,k_0$ and $y\,k_0$.  In both cases, we used $N=10$, $k_0=5$ and $\bar{V}=0.1$.}
\end{figure}

Fig.~\ref{fig:0a} shows a typical output of our numerical scheme for the scalar function $P$. For the $D=3$ numerics we show the scalar profile at the boundary as well as in the Lifshitz IR scaling regime. The latter is our disordered zero temperature horizon. The metric functions look equally disordered. Fig.~\ref{fig:0b} shows the result for a numerical simulation with $D=4$. In this case we show a density plot of the scalar field on the disordered IR Lifshitz horizon.

\begin{figure}[h]
\centering
\includegraphics[height = 0.3\textheight]{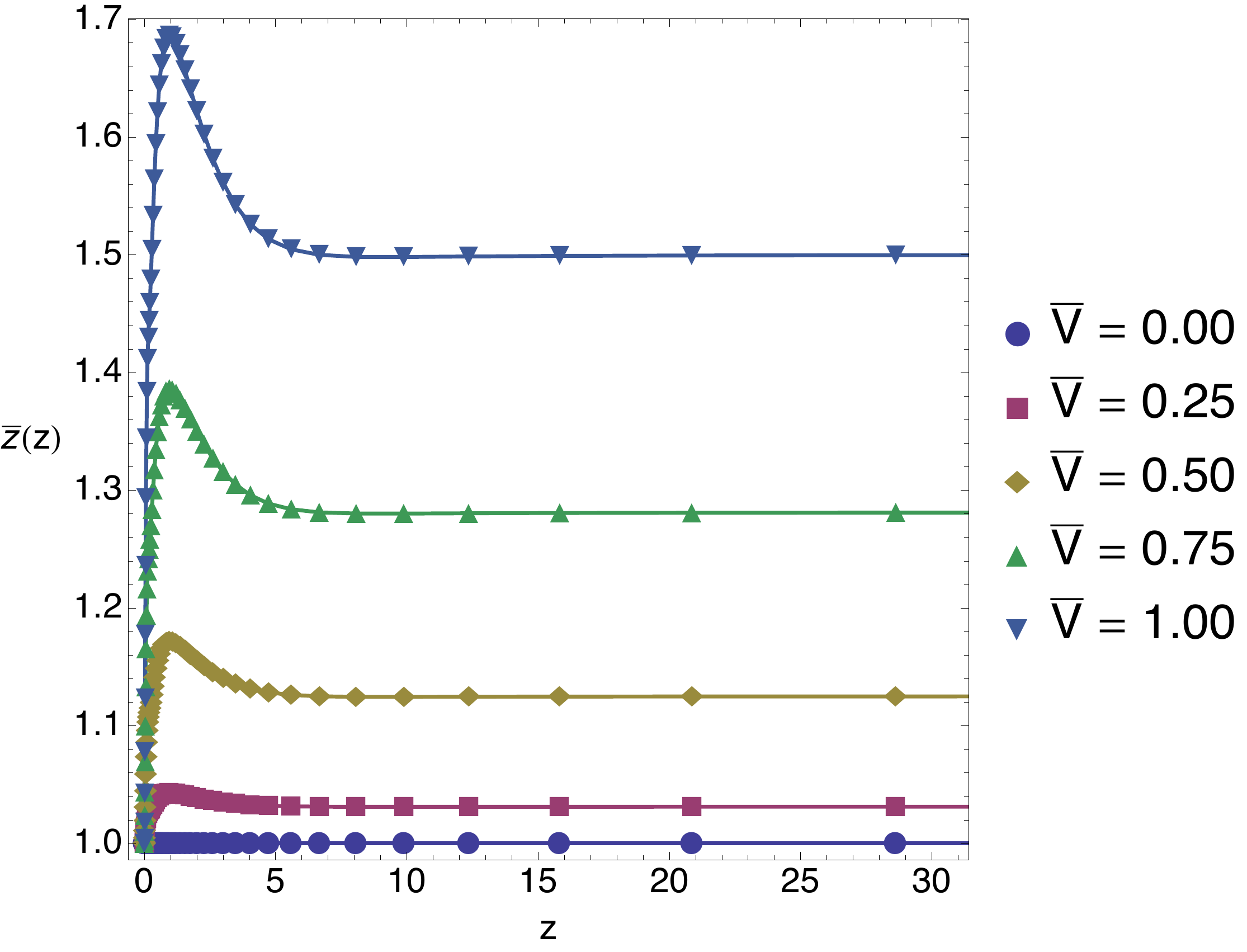}
\caption{\label{figs:1} {\bf Emergence of an IR dynamical scaling exponent}. Plot of ${\overline z}(z)$, the logarithmic derivative (\ref{eq:logderiv}) of $\langle g_{tt}\rangle_R$, for several values of $\bar{V}$. In the UV, ${\overline z} = 1$ for all curves. The constant values at large $z$ indicate an emergent Lifshitz scaling with ${\overline z}$ increasing with $\bar V$. These plots have $N=50$ and $k_0=5$.}
\end{figure}

Fig.~\ref{figs:1} shows several plots of the logarithmic derivative (\ref{eq:logderiv}) of $\langle g_{tt} \rangle_R$, as a function of the Fefferman-Graham coordinate $z$. It is clear that there is a wide range of large values of the coordinate $z$ over which there is scaling regime appropriate to extract ${\overline z}$. This regime does not extend all the way to $z \to \infty$ because our disorder has an IR cutoff at $N/k_0$. However, the scaling regime gets larger as $N$ increases and can be made parametrically large. We see in this plot that the existence of a scaling regime persists to larger values of $\bar V$ beyond the reach of perturbation theory.

\begin{figure}[h]
\centering
\subfigure{\label{fig:2} (a)\includegraphics[height = 0.27\textheight]{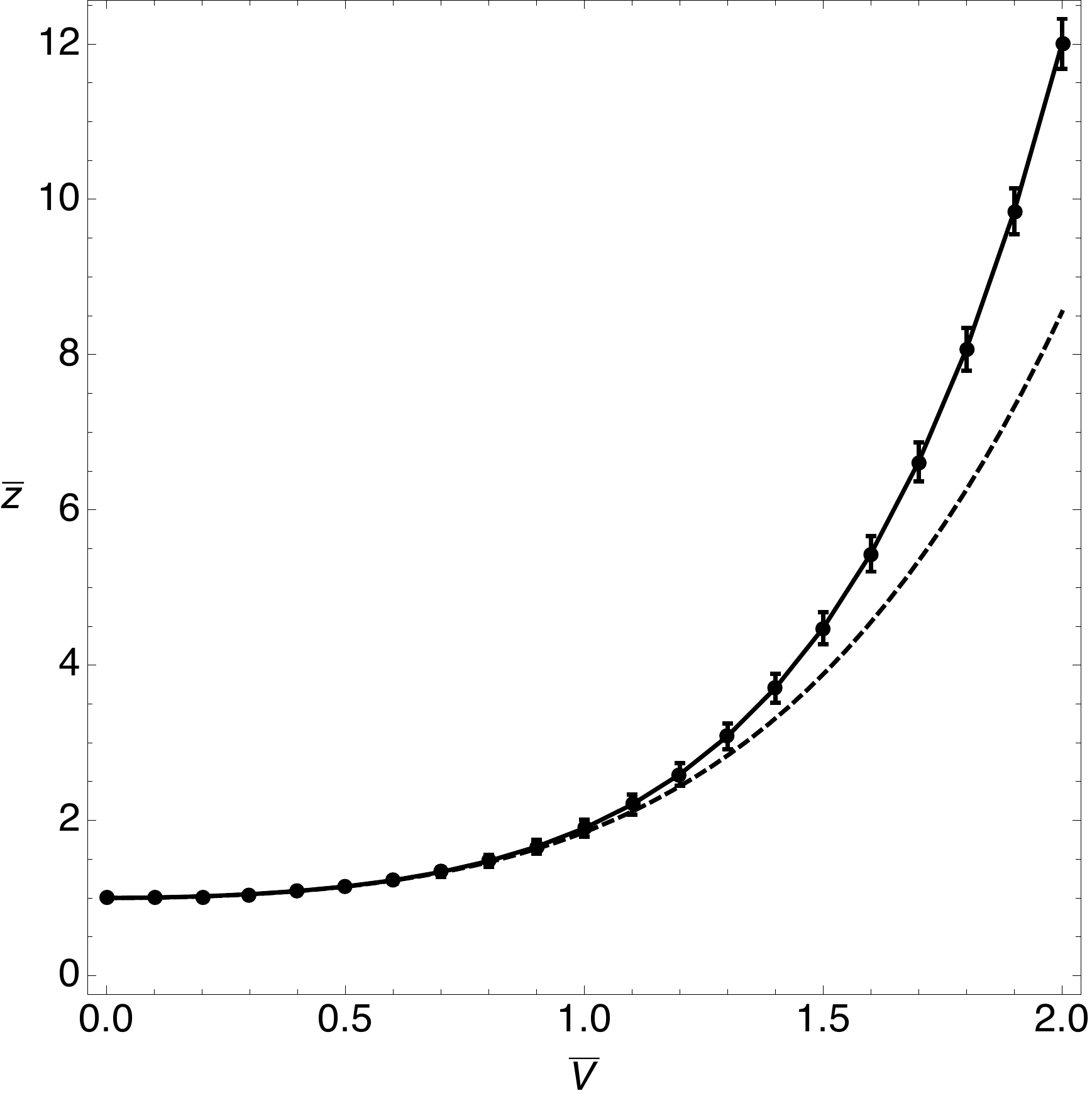}}
\subfigure{\label{fig:3} (b) \includegraphics[height = 0.27\textheight]{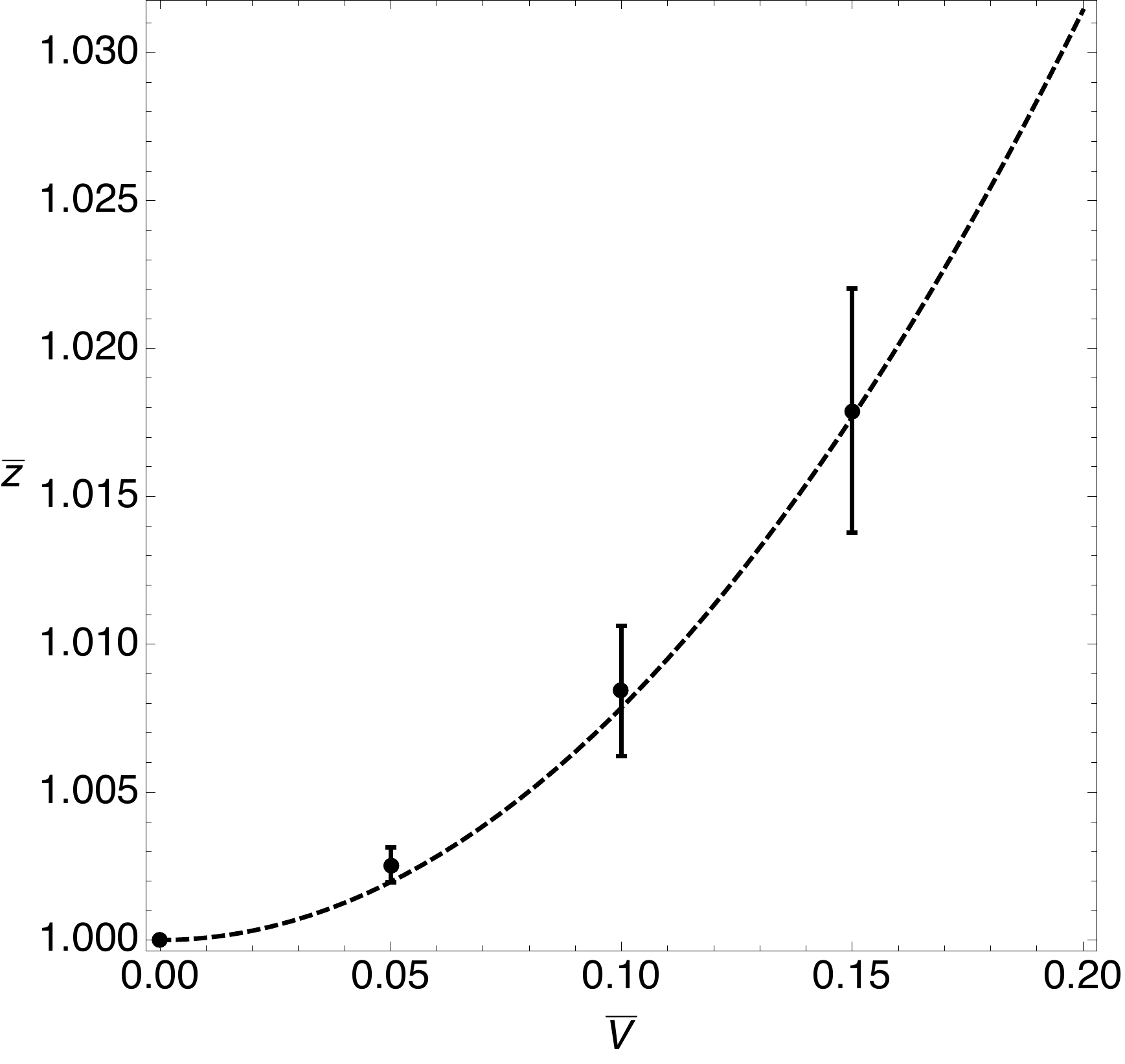}}
\caption{\label{figs:zb}{\bf Large disorder leads to large ${\overline z}$}: Dynamical critical exponent ${\overline z}$ as a function of $\bar{V}$. Plot (a) is for D=3 and plot (b) is for D=4. Numerical results are shown with statistical error bars while the dashed line is the perturbative expansion in $\bar{V}$. In $D=3$ we go well beyond the perturbative regime, while in $D=4$ we obtain a check of our analytic result.}
\end{figure}

Fig.~\ref{fig:2} shows the emergent scaling exponent ${\overline z}$ as a function of $\bar{V}$: the points represent our numerical data, each with its own error bar, and the dashed lines are the perturbative expansions (\ref{eq:zgen}) and (\ref{eq:zthree}). The error was extracted using a standard $\chi^2$ error. Of course, this does not take into account the discretization error, which is likely to be larger. It appears that the perturbative expansions capture the correct exponent until $\bar{V}\sim1$. Moreover, ${\overline z}$ seems to be a monotonically increasing function of $\bar{V}$, and we find no evidence of any saturation or other novel physics for the probed range of $\bar{V}$. Due to numerical lack of convergence, we did not manage to go beyond $\bar{V}=2$ (corresponding to ${\overline z} = 12$). The results suggest that ${\overline z}=+\infty$ can be reached if $\bar{V}\to+\infty$.

We also simulated disorder in $D=4$, \emph{i.e.} dealing with the full three-dimensional elliptic problem, but we did not succeeded in going beyond the perturbative regime, see Fig.~\ref{fig:3}. However, the numerical results seem to corroborate the perturbative expansion (\ref{eq:zgen}) in this more general setup.

\section{Discussion}

The objective of this paper has been the construction of zero temperature bulk geometries describing the deformation of a CFT by
a marginally relevant disordered coupling. We have found that the disorder drives the theory to a new disordered IR fixed point, characterized by a Lifshitz scaling with dynamical critical exponent $\overline z > 1$. Thus we have found a new class of spacetimes with zero temperature horizons.

Looking forward, it seems likely that these solutions will exhibit an interesting condensed matter phenomenology. As emphasized in the introduction, the primary effect of disorder is to break translation invariance. Because disorder involves modes with arbitrarily long wavelength, it is able to dissipate momentum efficiently in circumstances where a lattice is not. Efficient momentum relaxation by a lattice requires the existence of low energy excitations with finite momentum. This occurs with a Fermi surface or if $\overline z = \infty$ \cite{Hartnoll:2012rj,Horowitz:2012ky}, but is typically not possible at finite $\overline z$ \cite{Hartnoll:2012rj, Hartnoll:2012wm}. In contrast, weak disorder leads to a momentum relaxation rate with a power law dependence on temperature at finite $\overline z$ \cite{Hartnoll:2007ih, Hartnoll:2008hs, Hartnoll:2014gba, Lucas:2014zea}.

The systems we have studied in this paper are at zero charge density and therefore have a universal electrical conductivity even in the absence of disorder, due to particle-hole symmetry \cite{ph1,ph2}. However, at a nonzero temperature the energy current will couple to the momentum (in fact, because our UV theory is Lorentz invariant, the energy current is equal to the momentum) and hence the thermal conductivity will be divergent in the absence of disorder \cite{Mahajan:2013cja}. The finite effects of disorder in our IR geometry can be expected to lead to a nonzero momentum relaxation rate and hence a finite thermal conductivity.

An exciting open question is the fate of the IR geometry when the disorder becomes more strongly relevant. We hope to report on this question in the near future \cite{us}. We have suggested in the introduction that one possibility is that the system flows to a $\overline z = \infty$ fixed point. However, it is also conceivable that there is no meaningful averaged spacetime geometry in these cases and that instead one must directly describe a dramatically irregular `horizon'. For instance, there might be a complicated structure of metastable horizons as envisioned in \cite{Anninos:2013mfa}. Especially when combined with a finite charge density, the characterization of infinitely disordered fixed points in holography would open a path towards the study of localization physics in strongly interacting systems. New localization mechanisms may be found, analogous to those arising with relevant holographic lattices \cite{Donos:2012js}.

\section*{Acknowledgements}
 
This work is partially supported by a DOE Early Career Award, by a Sloan fellowship and by the Templeton Foundation.

\appendix

\section{Solution at second order in perturbation theory}
\label{sec:second}

This appendix contains the disordered solution for the metric at second order in perturbation theory. The solution to second order reads
\begin{subequations}
\begin{multline}
A^{(2)}(x,z) = G_1(x)+G_2(z)-z\sum_{j=0}^{N-1}A_j^2\,e^{-2k_j\,z}\cos^2(k_j\,x+\gamma_j)+\\
\sum_{i=1}^{N-1}\sum_{j=1}^{i-1}\frac{A_i\,A_j\,e^{-z (k_i+k_j)}}{\left(k_i+k_j\right)^3}\left[k_i^2 \left(1-4 z k_j\right)-2 k_i k_j \left(1+2 z k_j\right)+k_j^2\right]\cos[(k_i+k_j)x+\gamma_i+\gamma_j]+\\
\sum_{i=1}^{N-1}\sum_{j=1}^{i-1}\frac{A_i A_j e^{-z \left(k_i+k_j\right)}}{k_i+k_j}\cos[(k_i-k_j)x+\gamma_i-\gamma_j]
\end{multline}
\begin{multline}
B^{(2)}(x,z) = G_3(x)+z^2 G_4(x)+\sum_{j=0}^{N-1}\frac{A_j^2}{2\,k_j}\,e^{-2k_j\,z}\cos^2(k_j\,x+\gamma_j)-\sum_{j=1}^{N-1}\frac{A_j^2 e^{-2 z k_j} \left(1+2 z k_j\right)}{4 k_j}+\\
\sum_{i=1}^{N-1}\sum_{j=1}^{i-1}\frac{A_i A_j e^{-z \left(k_i+k_j\right)}}{k_i+k_j}\cos[(k_i+k_j)x+\gamma_i+\gamma_j]+\\
\sum_{i=1}^{N-1}\sum_{j=1}^{i-1}\frac{A_i\,A_j\,e^{-z (k_i+k_j)}}{\left(k_i+k_j\right)^3}\left[k_i^2 \left(1-4 z k_j\right)-2 k_i k_j \left(1+2 z k_j\right)+k_j^2\right]\cos[(k_i-k_j)x+\gamma_i-\gamma_j]\,,
\end{multline}
where $G_1(x)$ and $G_3(x)$ are integration functions, and
\begin{align}
&G_2(z) = \eta_1+z^2\,\eta_2+\sum_{j=1}^{N-1} A_j^2\,e^{-2\,k_j\,z}\frac{(1+z\,k_j)}{2\,k_j}\,,
\\
&G_4(x) = \frac{1}{2}\left[G^{\prime\prime}_1(x)-\eta_2\right]\,,
\end{align}
\label{eqs:aux}
\end{subequations}
with integration constants $\eta_1$ and $\eta_2$.

If we demand that the spacetime asymptotes to the line element (\ref{eq:boundarymetric}), this uniquely fixes $G_3(x)$. The condition\footnote{Note that upon a coordinate redefinition of the form $z\to p(x)z$, one can set the boundary metric to be exactly given by the line element (\ref{eq:boundarymetric}) if $p(x)$ is conveniently chosen.} comes from simply demanding that $A^{(2)}(x,0)=B^{(2)}(x,0)$. We are thus left with $\eta_1$, $\eta_2$ and $G_1(x)$ to be fixed by requiring regularity at the Poincar\'e horizon $z \to \infty$. This requirement sets $\eta_2 = G_1(x)=0$, so that $\eta_1$ is the only free constant, to be fixed at a later stage.

We are now ready to look at the disorder averaged metric functions, which we can easily read from Eqs.~(\ref{eqs:aux}):
\begin{align}
&\langle A^{(2)}(x,z) \rangle_{R} = \lim_{N\to+\infty}\left(\eta_1+ \sum_{j=1}^{N-1}\frac{e^{-2\,k_j\,z}A_j^2}{4\,k_j}\right)\,,\nonumber
\\
\\
&\langle B^{(2)}(x,z) \rangle_{R} =\lim_{N\to+\infty}\left[\eta_1+\sum_{j=1}^{N-1}\frac{A_j^2}{2}\,\left(\frac{1}{2\,k_j}-z\,e^{-2\,k_j\,z}\right)\right]\nonumber\,.
\end{align}
In our case of local Gaussian random noise ($A_i = \text{const.}$), we can actually do the sums in terms of elementary functions:
\begin{align}
\langle A^{(2)}(x,z) \rangle_{R} =\lim_{N\to+\infty}\left[\eta_1-e^{-2\,z\,k_0}G\left(e^{-\frac{2\,z\,k_0}{N}},1,N\right)-\ln\left(1-e^{-\frac{2\,z\,k_0}{N}}\right)\right]\,,\nonumber
\\
\\
\langle B^{(2)}(x,z) \rangle_{R} = \lim_{N\to+\infty}\left[\eta_1+H_{N-1}+2\,z\,k_0\,\frac{1-e^{-\frac{2\,z\,(N-1)\,k_0}{N}}}{N\left(1-e^{-\frac{2\,z\,k_0}{N}}\right)}\right]\,.\nonumber
\end{align}
where $G(a,b,c)$ is the Lerch transcendent function\footnote{The Lerch transcendent function, for $\mathrm{Re}(c)>0$, is defined as $$G(a,b,c) = \sum_{j=0}^{+\infty}\frac{a^k}{(k+c)^b} \,.$$} and $H_i$ the $i-$th harmonic number. The $N \to \infty$ limit can be taken easily, and provides the central result that we quoted in the main text:
\begin{align}
&\langle A^{(2)}(x,z) \rangle_{R} =e^{-2 k_0 z} \left\{(1+2 k_0 z) \left[\ln (2 k_0 z)+\gamma \right]-2 k_0 z\right\}-\ln (2 k_0 z)\,,\nonumber
\\
\label{eq:limitcentralappend}
\\
&\langle B^{(2)}(x,z)\rangle_{R} =e^{-2 k_0 z}+\gamma -1\,,\nonumber
\end{align}
where $\gamma$ is Euler's constant, and we have chosen $\eta_1 = -\ln N$, in order for the limit to be regular.

\section{\label{sec:generalD}Perturbation theory in general spacetime dimension}

In this appendix we outline the computation of the exponent (\ref{eq:genD}). Because we are after
the exponent to second order in $\bar{V}$, we only need to compute the averaged value of the second order energy momentum tensor.

Throughout this appendix, we keep the same notation as in Section \ref{sec:analytic}. The boundary spatial directions will be denoted $x_i$, for $i\in\{1,\ldots,D-2\}$.

We start with the first order solution. For $\Delta = D/2$, which corresponds to a scalar field saturating the Harris criterion, the linear solution reads
\begin{equation}
\widehat{\Phi}^{(1)}(x_1,\ldots,x_{D-2},z)=z^{\frac{D}{2}-1}\sum_{j_1,\ldots j_{D-2}=1}^{N-1}A_{j_1,\ldots,j_{D-2}}\,e^{-z\,K}\,\prod_{i=1}^{D-2} \cos(k_{i,j_i}\,x_i+\gamma_{i,j_i})\,,
\end{equation}
where each $\gamma_{i,j_i}$ is a random phase with a uniform distribution that take values in $(0,2\pi)$, and we have defined
\begin{equation}
K \equiv \sqrt{\sum_{m=1}^{D-2} k_{m,j_m}^2}\,.
\end{equation}
Note that for each of the $D-2$ boundary spatial dimensions, there is a different choice of wave-number distributions, corresponding to each of the $k_{i,j_i}$. We are now ready to present the averaged second order stress energy tensor. First, we note that because $\Phi$ is first order in $\bar{V}$, $\langle T^{(2)}_{ab}\rangle_R$ can be computed using the metric of pure AdS written in Poincar\'e coordinates. Second, we note that we are only interested in the second order averaged metric, which means that in this sector, to second order in $\bar{V}$ and in Fefferman-Graham coordinates, the metric is diagonal, and takes the following simple form:
\begin{equation}
\mathrm{d}s^2 = \frac{L^2}{z^2}\left[-(1+\bar{V}^2 \langle A^{(2)}\rangle_R)\mathrm{d}t^2+\sum_{i=1}^{D-2}(1+\bar{V}^2 \langle S_i^{(2)}\rangle_R)\mathrm{d}x_i^2+\mathrm{d}z^2\right]\,.
\label{eq:legeneralD}
\end{equation}
Note that $A^{(2)}$ itself is a function of all bulk spatial coordinates, but $\langle A^{(2)}\rangle_R$ is a function of $z$ only, by construction.

Setting $\bar{V}=0$ in the line element above gives the pure AdS metric needed to compute the averaged energy momentum tensor. This is easily done, and yields
\begin{subequations}
\begin{align}
&\theta_{tt}\equiv\langle \tilde{T}^{(2)}_{tt}\rangle^N_R = \frac{D\,z^{D-4}}{2^{D-1}}\,\sum_{j_1,\ldots j_{D-2}=1}^{N-1}A^2_{j_1,\ldots,j_{D-2}}\,e^{-2\,z\,K}\,,
\\
&\theta_{ii}\equiv\langle \tilde{T}^{(2)}_{ii}\rangle^N_R = -\frac{z^{D-4}}{2^{D-1}}\,\sum_{j_1,\ldots j_{D-2}=1}^{N-1}A^2_{j_1,\ldots,j_{D-2}}\,e^{-2\,z\,K}(D-4\,z^2\,k_{i,j_i}^2)\,,
\\
&\theta_{zz}\equiv\langle \tilde{T}^{(2)}_{zz}\rangle^N_R =-\frac{z^{D-4}}{2^{D-1}}\,\sum_{j_1,\ldots j_{D-2}=1}^{N-1}A^2_{j_1,\ldots,j_{D-2}}\,e^{-2\,z\,K}\left[D-z\,\left(D-2-2K\right)^2\right]\,,
\end{align}
\end{subequations}
where $\lim_{N\to+\infty}\langle\cdot\rangle^N_R = \langle\cdot\rangle_R$ and $\tilde{T}$ is the trace reversed energy momentum tensor
\begin{equation}
\tilde{T}_{ab}= T_{ab}-\frac{T}{D-2}g_{ab}\,.
\end{equation}
On the other hand, the perturbed and averaged Einstein equations read
 \begin{subequations}
 \begin{align}
&a^{\prime\prime}-\frac{D-1}{z}a^{\prime}-\frac{1}{z}\sum_{i=1}^{D-2}s^{\prime}_i-2\theta_{tt}=0\,,
\\
&s^{\prime\prime}_i-\frac{D-1}{z}s^{\prime}_i-\frac{a^\prime}{z}-\frac{1}{z}\sum_{j\neq i\atop j=1}^{D-2}s^{\prime}_j+2\theta_{ii}=0\,,
\\
&a^{\prime\prime}-\frac{a^{\prime}}{z}+\sum_{i=1}^{D-2}\left[s_i^{\prime\prime}-\frac{s_i^{\prime}}{z}\right]+2\theta_{zz}=0\,,
 \end{align}
\end{subequations}
 where we defined $a\equiv \langle A^{(2)}\rangle^N_R$ and $s_i \equiv \langle S_i^{(2)}\rangle^N_R$, for the sake of notation. These equations are readily solved as a function of $N$,
  \begin{subequations}
 \begin{align}
& a(z) = \frac{1}{2^{2(D-2)}}\sum_{j_1,\ldots j_{D-2}=1}^{N-1}\frac{A^2_{j_1,\ldots,j_{D-2}}}{K^{D-2}}\,\widetilde{\Gamma}\left(D-2,2z\,K\right)\,,
\\
& s_i(z) = \frac{1}{2^{2(D-2)}}\sum_{j_1,\ldots j_{D-2}=1}^{N-1}\frac{A^2_{j_1,\ldots,j_{D-2}}}{K^D}\,\Bigg[K^2\widetilde{\Gamma}\left(D-2,2z\,K\right)-k_{i,j_i}^2\,\widetilde{\Gamma}\left(D-1,2z\,K\right)\Bigg]\,,
 \end{align}
 \end{subequations}
 where $\widetilde{\Gamma}$ is the incomplete Gamma function and we fixed some of the integration constants so that, after a gauge transformation, at $z=0$ the bulk spacetime asymptotes to the line element (\ref{eq:legeneralD}) with $\bar{V}=0$.
 
 In order to finish the calculation we need to evaluate these sums. Here we shall only detail how to evaluate the first of the above two. First, we make a simple but useful remark. Namely, we can take a derivative of $a$, to find
 \begin{equation}
 a^{\prime}(z) = -\frac{1}{2^{D-2}}\sum_{j_1,\ldots j_{D-2}=1}^{N-1}A^2_{j_1,\ldots,j_{D-2}}\,z^{D-3}e^{-2\,K\,z}\,.
 \end{equation}
 It turns out this sum is much easier to evaluate, and after a simple integration in $z$ gives the result we desire. Second, if we are interested in local Gaussian disorder, we regard each dimension as having independent local Gaussian statistics, which in turn implies
 \begin{equation}
 A_{j_1,\ldots,j_{D-2}}=(2\,\sqrt{\Delta k})^{D-1}\,,\quad k_{i,j_i} = j_i\,\Delta k\quad\text{and}\quad\Delta k = k_0/N\,.
 \end{equation}
 We are thus left with evaluating the following sum
 \begin{equation}
 a^{\prime}(z) = -\frac{2^{D-2}\,k_0^{D-2}}{N^{D-2}}\sum_{j_1,\ldots j_{D-2}=1}^{N-1}\,z^{D-3}\exp\left[-2\,z\,\frac{k_0}{N}\sqrt{\sum_{m=1}^{D-2}j_m^2}\right]\,,
 \end{equation}
 which, in the limit $N\to+\infty$, we recognize as a multi-dimensional Riemann sum on the multi-dimensional hypercube $(0,1)^{D-2}$, \emph{i.e.}
 \begin{equation}
 \langle A^{(2)}\rangle_R^{\prime}(z)=-\frac{2^{D-2}}{z}\tilde{k}^{D-2}\prod_{m=1}^{D-2}\int_0^1\mathrm{d}x_m\,\exp\left(-2\tilde{k}\sqrt{\sum_{m=1}^{D-2}x_m^2}\right)\,,
 \label{eq:interm}
 \end{equation}
 where we defined $\tilde{k} = k_0\,z$. This is our final expression if we are interested in knowing the averaged $g_{tt}$, to second order in $\bar{V}$, and for $z\in(0,+\infty)$. If, however, we are interested in the behavior of this function as $z\to+\infty$, we can make further progress. First, we make the following change of coordinates
 \begin{equation}
y_m = \tilde{k}\,x_m\,,
 \end{equation}
 which brings Eq.~(\ref{eq:interm}) to:
 \begin{equation}
\langle A^{(2)}\rangle_R^{\prime}(z)=-\frac{2^{D-2}}{z}\prod_{m=1}^{D-2}\int_0^{\tilde{k}}\mathrm{d}y_m\,\exp\left(-2\sqrt{\sum_{m=1}^{D-2}y_m^2}\right)\,.
 \end{equation}
If $z$ is very large, then so is $\tilde{k}$, in which case the integral above can be easily done in terms of complete Gamma functions:
 \bea
\lefteqn{\langle A^{(2)}\rangle_R^{\prime}(z)= -\frac{\pi ^{\frac{D-3}{2}}}{z} \Gamma \left(\frac{D-1}{2}\right)+\mathcal{O}(e^{-2 k_0 z})} \nonumber\\ && \Rightarrow \quad \langle A^{(2)}\rangle_R(z) = -\pi ^{\frac{D-3}{2}} \Gamma \left(\frac{D-1}{2}\right)\ln (z\,k_0)+\mathcal{O}(e^{-2 k_0 z})\,.
 \eea
This is the logarithmic divergence that we resum to obtain Eq.~(\ref{eq:genD}).


\begin{thebibliography}{99}
 
 \bibitem{harris}
A.~B.~Harris, ``Effect of random defects on the critical behaviour of Ising models,''
 J. of Phys. C {\bf 7}, 1671 (1974).
 
\bibitem{Weinrib:1983zz} 
  A.~Weinrib and B.~I.~Halperin,
  ``Critical phenomena in systems with long-range-correlated quenched disorder,''
  Phys.\ Rev.\ B {\bf 27}, 413 (1983).

\bibitem{sachdev}
S.~Sachdev, {\it Quantum phase transitions}, CUP 1999.

\bibitem{mot}
O.~Motrunich, S.-C.~Mau, D.~A.~Huse, D.~S.~Fisher,
``Infinite-randomness quantum Ising critical fixed points,''
Phys. Rev. B {\bf 61}, 1160 (2000)
[arXiv:cond-mat/9906322 [cond-mat.dis-nn]].

\bibitem{vojta}
T.~Vojta, ``Rare region effects at classical, quantum and nonequilibrium phase transitions,''
J. Phys. A {\bf 39} R143 (2006)
[arXiv:cond-mat/0602312 [cond-mat.stat-mech]].

\bibitem{Hartnoll:2007ih} 
  S.~A.~Hartnoll, P.~K.~Kovtun, M.~Muller and S.~Sachdev,
  ``Theory of the Nernst effect near quantum phase transitions in condensed matter, and in dyonic black holes,''
  Phys.\ Rev.\ B {\bf 76}, 144502 (2007)
  [arXiv:0706.3215 [cond-mat.str-el]].

\bibitem{Hartnoll:2008hs} 
  S.~A.~Hartnoll and C.~P.~Herzog,
  ``Impure AdS/CFT correspondence,''
  Phys.\ Rev.\ D {\bf 77}, 106009 (2008)
  [arXiv:0801.1693 [hep-th]].
  
\bibitem{Hartnoll:2012rj} 
  S.~A.~Hartnoll and D.~M.~Hofman,
  ``Locally Critical Resistivities from Umklapp Scattering,''
  Phys.\ Rev.\ Lett.\  {\bf 108}, 241601 (2012)
  [arXiv:1201.3917 [hep-th]].
  
\bibitem{Anantua:2012nj} 
  R.~J.~Anantua, S.~A.~Hartnoll, V.~L.~Martin and D.~M.~Ramirez,
  ``The Pauli exclusion principle at strong coupling: Holographic matter and momentum space,''
  JHEP {\bf 1303}, 104 (2013)
  [arXiv:1210.1590 [hep-th]].
  
\bibitem{Lucas:2014zea} 
  A.~Lucas, S.~Sachdev and K.~Schalm,
  ``Scale-invariant hyperscaling-violating holographic theories and the resistivity of strange metals with random-field disorder,''
  arXiv:1401.7993 [hep-th].
  
\bibitem{Adams:2011rj} 
  A.~Adams and S.~Yaida,
  ``Disordered Holographic Systems I: Functional Renormalization,''
  arXiv:1102.2892 [hep-th].

\bibitem{Adams:2012yi} 
  A.~Adams and S.~Yaida,
  ``Disordered Holographic Systems II: Marginal Relevance of Imperfection,''
  arXiv:1201.6366 [hep-th].
  
\bibitem{Kachru:2008yh} 
  S.~Kachru, X.~Liu and M.~Mulligan,
  ``Gravity Duals of Lifshitz-like Fixed Points,''
  Phys.\ Rev.\ D {\bf 78}, 106005 (2008)
  [arXiv:0808.1725 [hep-th]].
  
  \bibitem{us}
  {\it Work in progress.}
  
  \bibitem{danfisher}
  D.~S.~Fisher,
  ``Random transverse field Ising spin chains,''
  Phys. Rev. Lett. {\bf 69}, 534 (1992).
  
 \bibitem{ykwen} 
 Y.~B.~Kim and X.-G.~Wen,
  ``Large-N renormalization-group study of the commensurate dirty-boson problem,''
   Phys.\ Rev.\ B {\bf 49}, 4043 (1994).
   
 \bibitem{cardy}
 D.~Boyanovsky and J.~L.~Cardy
``Critical behavior of m-component magnets with correlated impurities,''
Phys. Rev. B {\bf 26}, 154 (1982).
  
\bibitem{Maldacena:1997re} 
  J.~M.~Maldacena,
  ``The Large N limit of superconformal field theories and supergravity,''
  Adv.\ Theor.\ Math.\ Phys.\  {\bf 2}, 231 (1998)
  [hep-th/9711200].
  
\bibitem{Witten:1998qj} 
  E.~Witten,
  ``Anti-de Sitter space and holography,''
  Adv.\ Theor.\ Math.\ Phys.\  {\bf 2}, 253 (1998)
  [hep-th/9802150].
  
\bibitem{Gubser:1998bc} 
  S.~S.~Gubser, I.~R.~Klebanov and A.~M.~Polyakov,
  ``Gauge theory correlators from noncritical string theory,''
  Phys.\ Lett.\ B {\bf 428}, 105 (1998)
  [hep-th/9802109].
  
\bibitem{Shinozuka1991}  
  M.~Shinozuka and G.~Deodatis, ``Simulation of stochastic processes by spectral representation,'' Appl. Mech. Rev. {\bf 44} (1991) 191.
 
\bibitem{Arean:2013mta} 
  D.~Arean, A.~Farahi, L.~A.~Pando Zayas, I.~S.~Landea and A.~Scardicchio,
  ``A Dirty Holographic Superconductor,''
  arXiv:1308.1920 [hep-th].
  
\bibitem{Dias:2011ss} 
  O.~J.~C.~Dias, G.~T.~Horowitz and J.~E.~Santos,
  ``Gravitational Turbulent Instability of Anti-de Sitter Space,''
  Class.\ Quant.\ Grav.\  {\bf 29}, 194002 (2012)
  [arXiv:1109.1825 [hep-th]].
  
\bibitem{Dias:2012tq} 
  O.~J.~C.~Dias, G.~T.~Horowitz, D.~Marolf and J.~E.~Santos,
  ``On the Nonlinear Stability of Asymptotically Anti-de Sitter Solutions,''
  Class.\ Quant.\ Grav.\  {\bf 29}, 235019 (2012)
  [arXiv:1208.5772 [gr-qc]].
  
\bibitem{Horowitz:2012ky} 
  G.~T.~Horowitz, J.~E.~Santos and D.~Tong,
  ``Optical Conductivity with Holographic Lattices,''
  JHEP {\bf 1207}, 168 (2012)
  [arXiv:1204.0519 [hep-th]].
  
\bibitem{Headrick:2009pv}
  M.~Headrick, S.~Kitchen and T.~Wiseman,
  ``A New approach to static numerical relativity, and its application to Kaluza-Klein black holes,''
  Class.\ Quant.\ Grav.\  {\bf 27} (2010) 035002
  [arXiv:0905.1822 [gr-qc]].
  
\bibitem{Figueras:2011va} 
  P.~Figueras, J.~Lucietti and T.~Wiseman,
  ``Ricci solitons, Ricci flow, and strongly coupled CFT in the Schwarzschild Unruh or Boulware vacua,''
  Class.\ Quant.\ Grav.\  {\bf 28}, 215018 (2011)
  [arXiv:1104.4489 [hep-th]].
   
\bibitem{Hartnoll:2012wm} 
  S.~A.~Hartnoll and E.~Shaghoulian,
  ``Spectral weight in holographic scaling geometries,''
  JHEP {\bf 1207}, 078 (2012)
  [arXiv:1203.4236 [hep-th]].
  
\bibitem{Hartnoll:2014gba} 
  S.~A.~Hartnoll, R.~Mahajan, M.~Punk and S.~Sachdev,
  ``Transport near the Ising-nematic quantum critical point of metals in two dimensions,''
  arXiv:1401.7012 [cond-mat.str-el].
    
  \bibitem{ph1}
  M.~P.~A.~Fisher, G.~Grinstein and S.~M.~Girvin,
  ``Presence of quantum diffusion in two dimensions: Universal resistance at the superconductor-insulator transition,''
  Phys. Rev. Lett. {\bf 64}, 587 (1990).
  
  \bibitem{ph2}
  K.~Damle and S.~Sachdev,
  ``Nonzero-temperature transport near quantum critical points,''
  Phys. Rev. B {\bf 56}, 8714 (1997).
  
\bibitem{Mahajan:2013cja} 
  R.~Mahajan, M.~Barkeshli and S.~A.~Hartnoll,
  ``Non-Fermi liquids and the Wiedemann-Franz law,''
  Phys.\ Rev.\ B {\bf 88}, 125107 (2013)
  [arXiv:1304.4249 [cond-mat.str-el]].
  
\bibitem{Anninos:2013mfa} 
  D.~Anninos, T.~Anous, F.~Denef and L.~Peeters,
  ``Holographic Vitrification,''
  arXiv:1309.0146 [hep-th].
  
\bibitem{Donos:2012js} 
  A.~Donos and S.~A.~Hartnoll,
  ``Interaction-driven localization in holography,''
  Nature Phys.\  {\bf 9}, 649 (2013)
  [arXiv:1212.2998].
  
  \end{thebibliography}
\end{document}